%% file: main.tex
\documentclass[10pt]{IEEEtran}
\usepackage{amsmath}
\usepackage{amsfonts}
\usepackage{amssymb}
\usepackage[ruled,vlined]{algorithm2e}
\usepackage{caption}
\usepackage{mathtools}  
\usepackage{amsthm}
\usepackage[]{algorithm2e}
\usepackage{booktabs}
\usepackage{subfigure}
\usepackage{tabulary}
\usepackage{footnote}
\usepackage[export]{adjustbox}
\usepackage{booktabs}
\usepackage[justification=centering]{caption}
\usepackage{comment}

\usepackage{cases}
\usepackage{graphicx}
\usepackage{cite}
\usepackage{multicol}
\usepackage{xcolor}
\usepackage{graphicx}
\graphicspath{{./}{figures/}}
\newcommand\numeq[1]%
  {\stackrel{\scriptscriptstyle(\mkern-1.5mu#1\mkern-1.5mu)}{=}}
\usepackage{stfloats}
\usepackage{cuted}
\usepackage{lipsum}  

\usepackage{subfigure}
\usepackage[nolist,nohyperlinks]{acronym}

\title{Comparison of STR and EMLSR Performance in Wi-Fi 7 MLO}

\author{\text{Aishwarya
 Choorakuzhiyil},  \text{Kevin Ho}, \text{and Sara Reyes}}%

\begin{document}

\maketitle

\begin{abstract}
This project compares the performance of simultaneous transmit and receive (STR) and enhanced multi-link single radio (EMLSR) within Multi-Link Operation (MLO) in Wi-Fi 7 networks. Using the ns-3 simulator, we evaluate both techniques under various scenarios, including changes in modulation coding scheme (MCS), bandwidth, link quality, and interference levels. Key performance metrics such as latency, throughput, and energy efficiency are analyzed to determine the trade-offs between STR and EMLSR. The results demonstrate that STR achieves higher throughput and lower latency due to dual-link utilization, making it suitable for high-load environments. In contrast, EMLSR balances energy efficiency with responsiveness, making it advantageous for power-sensitive applications. This analysis provides insights into the strengths and limitations of STR and EMLSR, guiding optimal deployment strategies for future Wi-Fi 7 networks.
\end{abstract}

\input{introduction}
\input{sys_model}
\input{simulation}

\input{conclusion}



\bibliographystyle{IEEEtran}
\bibliography{reference}
\end{document}

%% file: introduction.tex
\section{Introduction} 
\label{introduction}

In this paper, we introduce a comparison of simultaneous transmit and receive (STR) mode and enhanced multi-link single radio (EMLSR) mode within Multi-Link Operation (MLO). We compare these two methods against each other and against Single-Link Operation (SLO). The IEEE 802.11be standard, known colloquially as Wi-Fi 7, introduced MLO, which enables devices to simultaneously send and receive data across different frequency bands and channels. For older generations of Wi-Fi, like Wi-Fi 5 or 6 only one band was available to the client. In addition to enabling MLO, Wi-Fi 7 also introduced channel bandwidths of 320~MHz which can deliver increased throughput gains \cite{10.1145/3659111.3659116}.

STR enables simultaneous asynchronous transmission and reception across multiple frequency bands, and is designed to maximize throughput, minimize latency, and enhance network performance. Meanwhile EMLSR, the other mode within MLO, enables listening on multiple links simultaneously while transmitting on only one, optimizing responsiveness and energy efficiency. It is also designed to balance power efficiency and multi-link capabilities, making it ideal for power-sensitive applications \cite{10.1145/3659111.3659116}. Previous works  have focused on analyzing either only STR mode \cite{10.1145/3659111.3659116, 10624802, 10495351} or EMLSR mode \cite{https://doi.org/10.1155/2022/7018360}. This paper aims to compare both modes of operation against each other as well as against SLO of older generations of Wi-Fi.

%% file: sys_model.tex
\section{System Model}
\label{sec:sys_model}

The network behavior was simulated using ns-3, an open source, discrete-event network simulator. Version 3.41 of the software, developed by twenty-three authors, introduced key features of Wi-Fi 7 extremely high throughput (EHT) \cite{v3.41-release}. Crucially, this release augmented ns-3's native EMLSR support, while previous versions already introduced STR capabilities.

Custom \texttt{C++} scripts were developed to simulate the desired MLO-capable networks. Both STR and EMLSR modes were investigated under similar network configurations. The \texttt{C++} scripts introduced modifiable parameters to allow network configurations to be changed without the need to rebuild the source each time. Python scripts were leveraged to alter the network setups. The \texttt{C++} source code can be found in the \texttt{examples/} directory, while the Python scripts are located within the \texttt{experiments/} directory of the project's GitHub repository \cite{group5-github}.

%% file: simulation.tex
\section{Simulation} \label{simulation}
We constructed several scenarios in ns-3 to evaluate and compare STR and EMLSR: (1) base simulation, (2) throughput vs. network size, (3) MLO performance with varied MCS, (4) MLO performance with varied BW, and (5) MLO performance under interference. The first two experiments focus on evaluating the benefits offered by these two MLO modes over the SLO of older Wi-Fi generations. Meanwhile, the last three scenarios concentrate on specifically comparing STR against EMLSR. 

\subsection{SLO vs STR vs EMLSR - Base Simulation}
For the base simulation, we evaluated the network with one AP and five STAs with a payload of 1500 packets for SLO, STR, and EMLSR. This section delineates the throughput and latency performance metrics for the respective modes of operation and it serves as a reference comparison.

\begin{figure}[!h]
    \centering
    \includegraphics[width=0.8\linewidth]{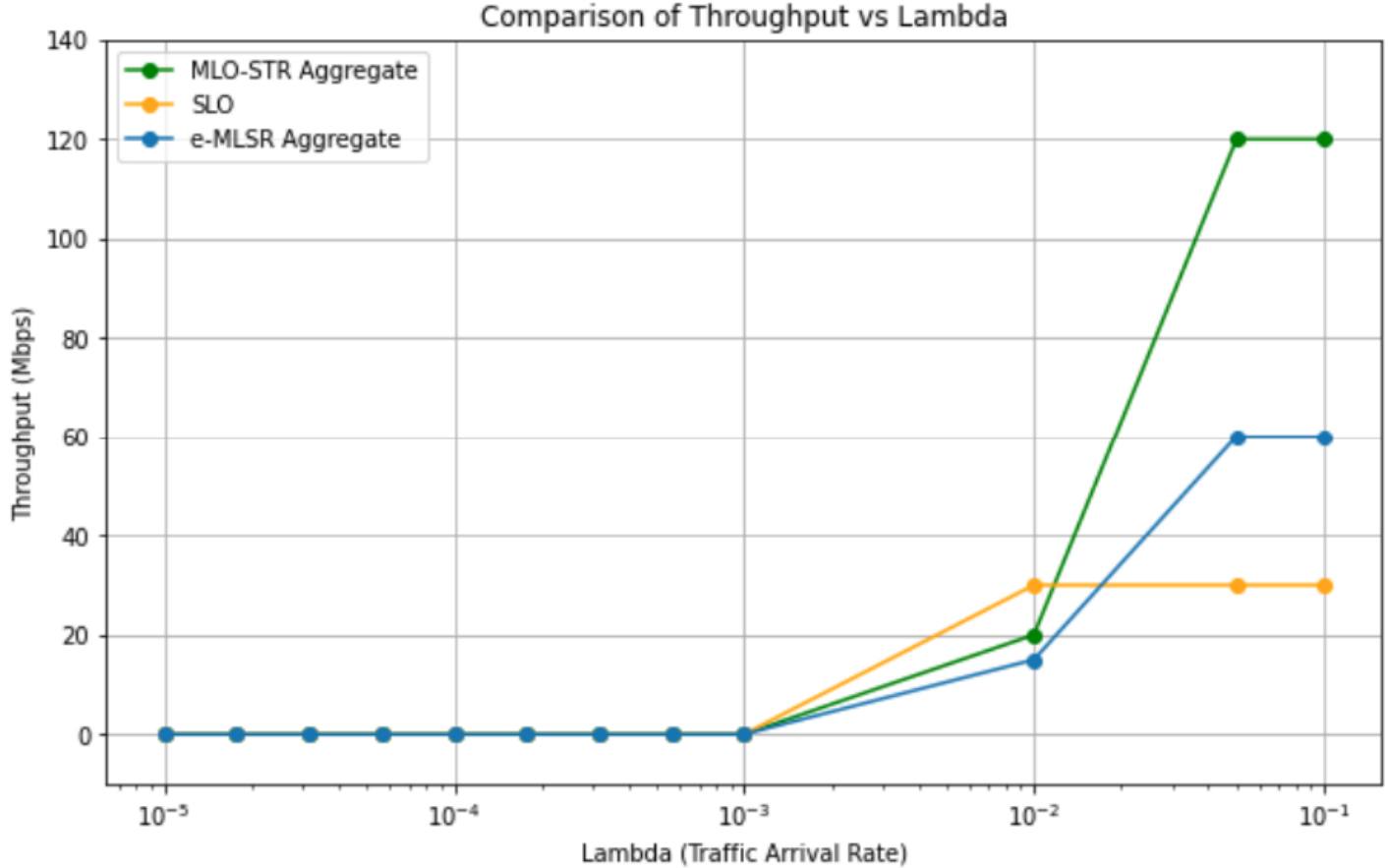}
    \caption{Throughput Comparison for SLO, STR, and EMLSR}
    \label{fig:throughput_comparison}
\end{figure}

Fig.~\ref{fig:throughput_comparison} shows the throughput vs. lambda comparison for SLO, STR, and EMLSR for five stations. It is observed that upon increasing the offload, the throughput obtained is the highest in STR. It is due to better traffic allocation and multi-link utilization contributing to a higher aggregate overall, thereby achieving a saturation point of about 120~Mbps at around $\lambda = 10^{-1}$. SLO has the least throughput due to its reliance on a single link of operation \cite{alsakati2023performance80211bewifi7} leading to increased congestion and saturation ~ at about 30~Mbps at around $\lambda = 10^{-2}$. The performance of EMLSR falls in between as it dynamically selects one link at a time, saturating at about 60~Mbps at around $\lambda = 10^{-2}$. It can be inferred that while EMLSR has a better performance relative to SLO by balancing loads across links, its performance is limited by the single radio, hence achieving a lower throughput than STR.

\begin{figure}[!tbp]
    \centering
    \includegraphics[width=0.8\linewidth]{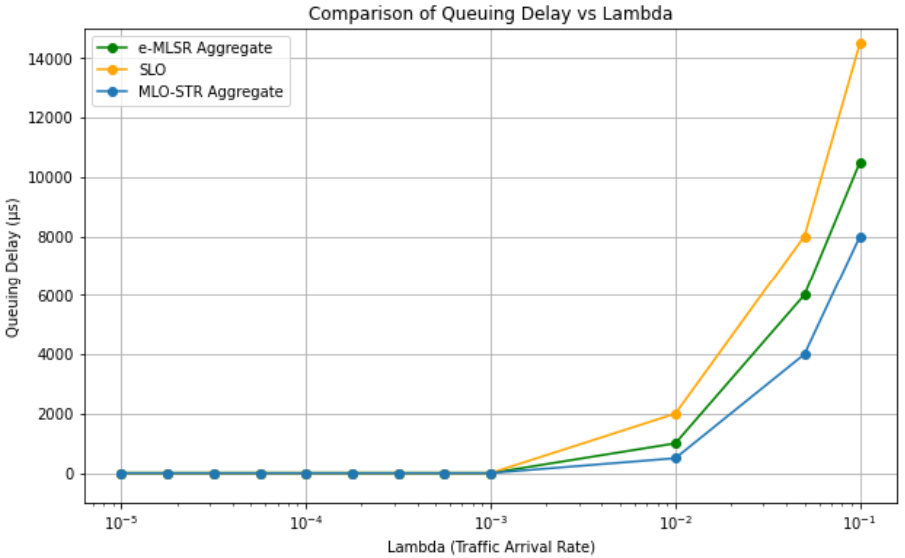}
    \caption{Queuing Delay Comparison for SLO, STR, and EMLSR}
    \label{fig:queuing_delay_comparison}
\end{figure}

\begin{figure}[!tbp]
    \centering
    \includegraphics[width=0.8\linewidth]{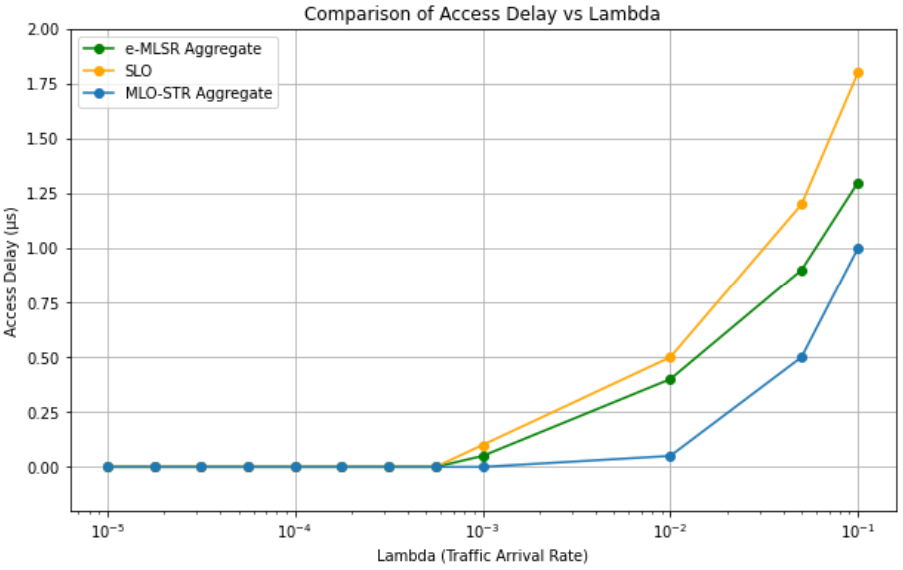}
    \caption{Access Delay Comparison for SLO, STR, and EMLSR}
    \label{fig:access_delay_comparison}
\end{figure}

\begin{figure}[!tbp]
    \centering
    \includegraphics[width=0.8\linewidth]{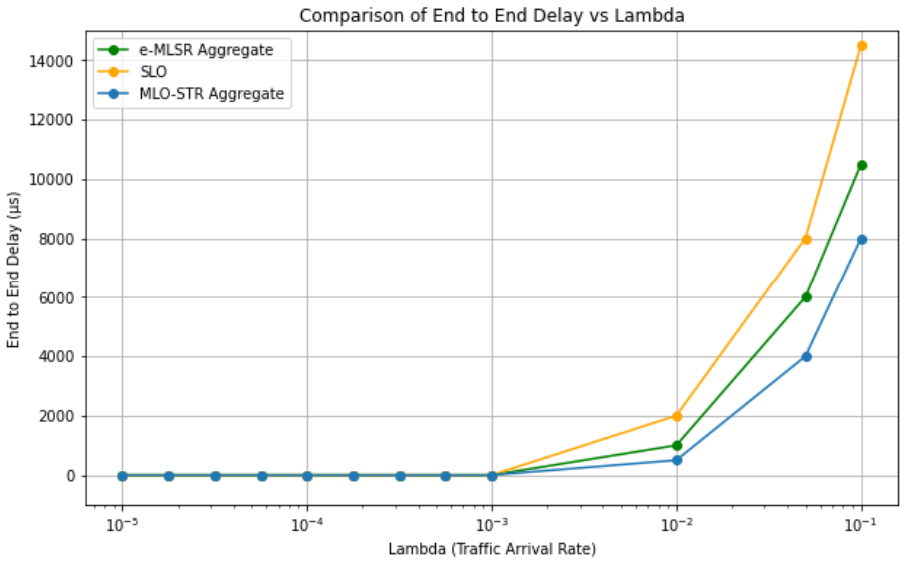}
    \caption{End to End Delay Comparison for SLO, STR, and EMLSR}
    \label{fig:e2e_delay_comparison}
\end{figure}

Queuing delay is the time taken for a packet to wait in a queue until the channel is accessible for transmission. Access delay is the time taken for a packet to access the channel for transmission, including the waiting time for the channel to become idle, back-off, contention process etc. End to end delay is the total time taken for packet transmission from source to destination, including all the delays the packet has encountered until it is received. 

Fig.~\ref{fig:queuing_delay_comparison}, Fig.~\ref{fig:access_delay_comparison} and Fig.~\ref{fig:e2e_delay_comparison} depict similar trends of latency with STR outperforming SLO and EMLSR, giving the least delays. The presence of only a single link of operation and poor traffic allocation leads to increased queuing, contention for access, high congestion; thereby causing significant delays in SLO especially under high loads. STR minimizes delays by dual link operation and balancing load, allowing packets to be transmitted faster with lower waiting times. EMLSR performs better than SLO by allocating load dynamically across links, but its single-radio constraint leads to higher waiting times than STR. 

\subsection{SLO vs STR vs EMLSR - Throughput vs Network Size}
For this simulation we evaluated the network with one AP and \{5, 10, ... , 30\} STAs, with a payload of 1500 packets.

\begin{figure}[!tbp]
    \centering
    \includegraphics[width=0.75\linewidth]{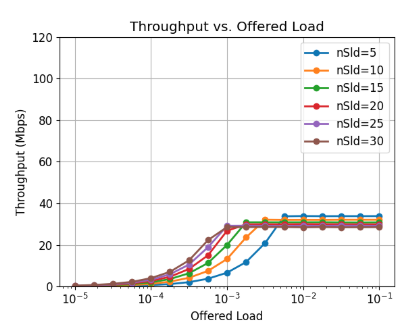}
    \caption{SLO: Throughput vs. Network Size}
    \label{fig:Network Size - SLO}
\end{figure}

Fig.~\ref{fig:Network Size - SLO} shows the throughout vs. offered load as the network size increases by increasing the number of stations simulated for SLO. The throughput seems to be saturated between 28 and 35~Mbps regardless of the number of stations. The lower the number of stations, the throughput saturates around $\lambda = 10^{-2}$, but with higher stations (i.e 15-30) the throughput starts to saturate earlier at $\lambda = 10^{-3}$. 

\begin{figure}[!tbp]
    \centering
    \includegraphics[width=0.75\linewidth]{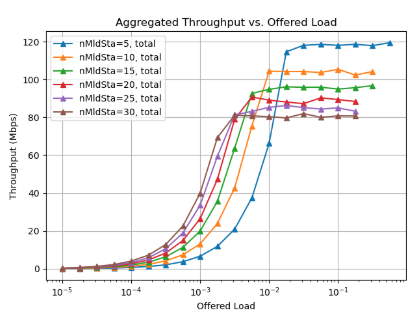}
    \caption{STR: Throughput vs. Network Size}
    \label{fig:Network Size - STR}
\end{figure}

Fig.~\ref{fig:Network Size - STR} shows the throughout vs. offered load as the network size increases by increasing the number of stations simulated for STR. With five STAs the throughput saturates at 120~Mbps, while with 10 STAs the throughput saturates at around 105~Mbps. For even higher number of STAs the throughput is saturating between 80 and 95~Mbps. As seen with the SLO plot, as the number of stations increases, the throughput saturates at a lower lambda value. 

\begin{figure}[!tbp]
    \centering
    \includegraphics[width=0.75\linewidth]{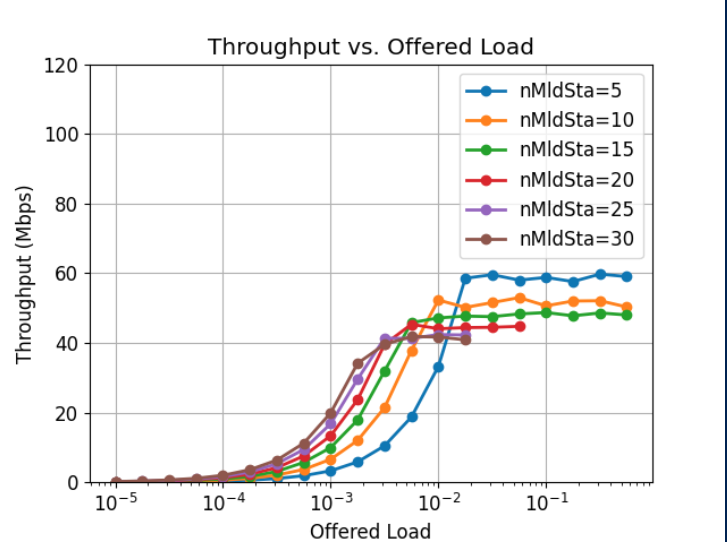}
    \caption{EMLSR: Throughput vs. Network Size}
    \label{fig:Network Size - EMLSR}
\end{figure}

Fig.~\ref{fig:Network Size - EMLSR} shows the throughout vs. offered load as the network size increases by increasing the number of stations simulated for EMLSR. The plot shows that for all STA's simulated the throughput saturates between 40 and 60~Mbps. The same pattern arises of the higher the amount of STA's the lower the lambda value is when the throughput saturates.  For EMLSR, the throughput saturation values are generally higher than when using SLO, but much lower than for STR mode. 

When comparing SLO vs. STR vs. EMLSR for Throughput vs. Network size. We can see that SLO performs the worst out of all three scenarios, which aligns with previous theory \cite{10149044}. Even as network size increases MLO will attain higher throughput gains, and MLO will outperform SLO.

\subsection{STR vs EMLSR - Varied MCS}\label{subsec:STR vs EMSR - Varied MCS}
\begin{figure}[!tbp]
    \centering
    \includegraphics[width=0.95\linewidth]{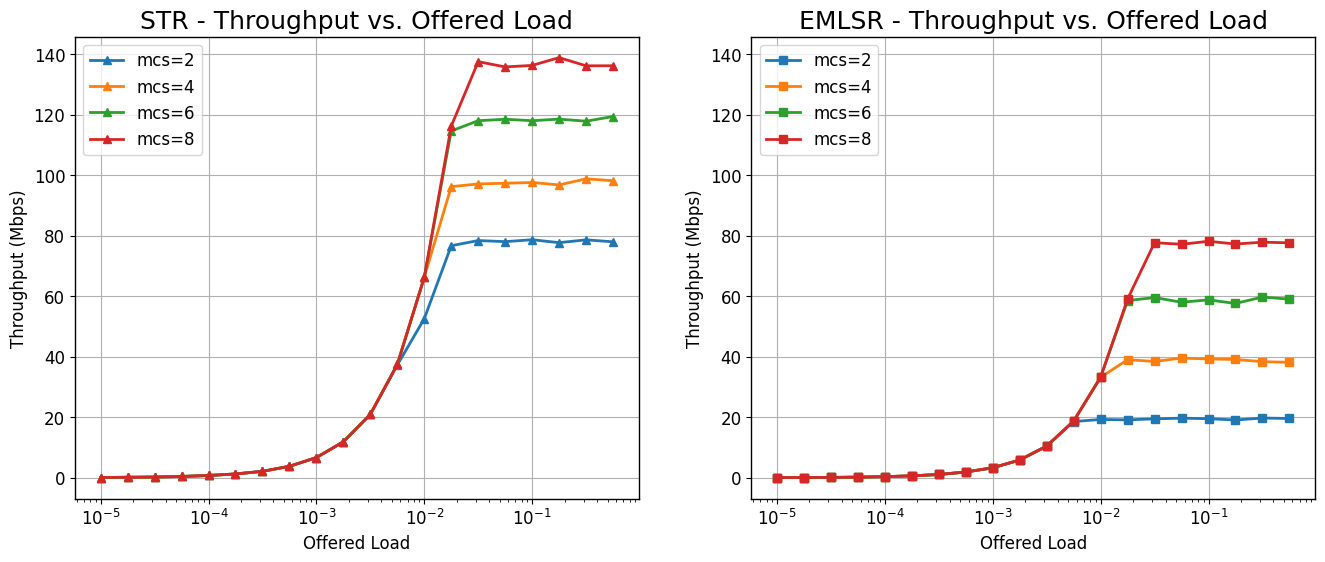}
    \caption{STR vs EMLSR: Throughput vs Offered Load with varied MCS}
    \label{fig:Varied MCS - T vs OL}
\end{figure}

\begin{figure}[!tbp]
    \centering
    \includegraphics[width=0.95\linewidth]{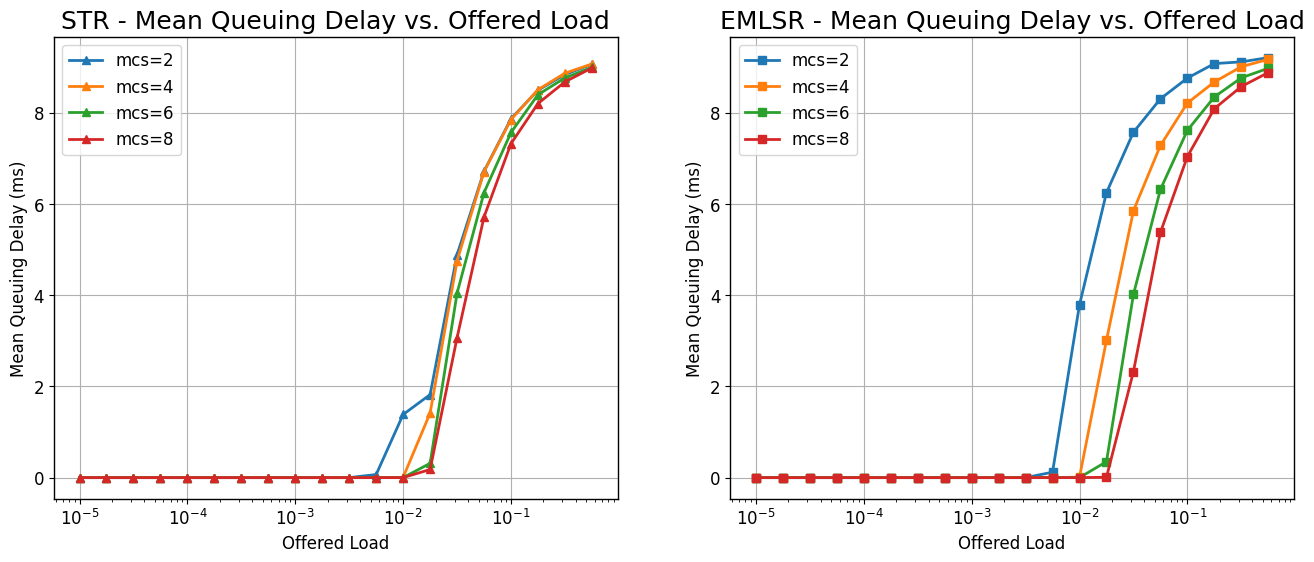}
    \caption{STR vs EMLSR: Mean Queuing Delay vs Offered Load with varied MCS}
    \label{fig:Varied MCS - QD vs OL}
\end{figure}

\begin{figure}[!tbp]
    \centering
    \includegraphics[width=0.95\linewidth]{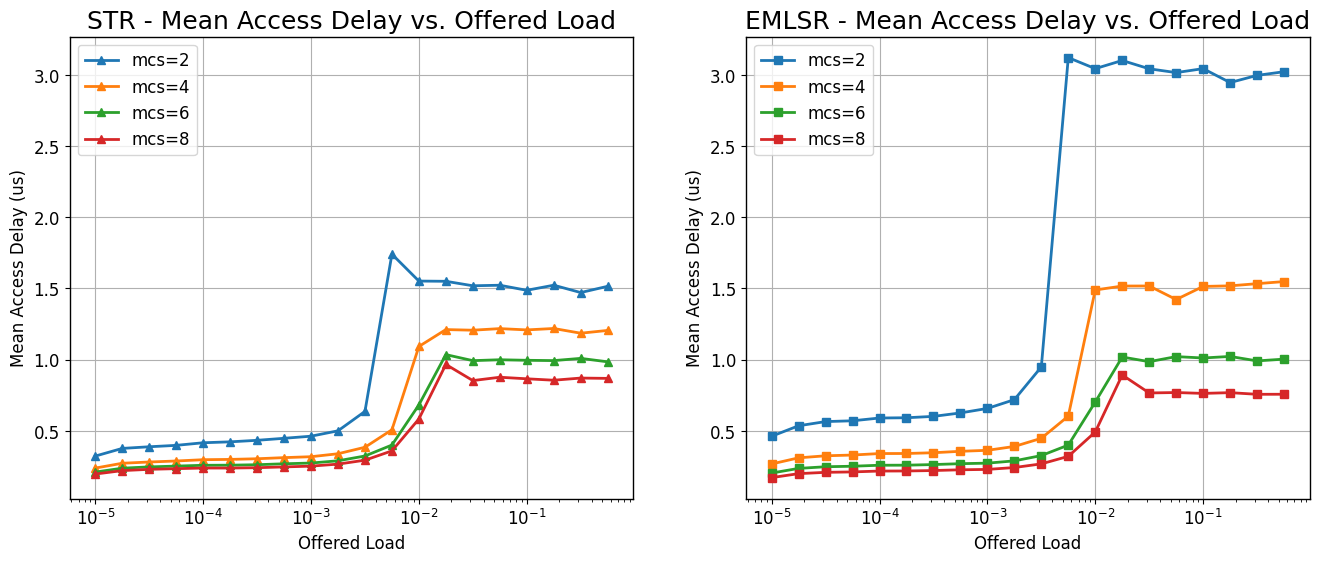}
    \caption{STR vs EMLSR: Mean Access Delay vs Offered Load with varied MCS}
    \label{fig:Varied MCS - AD vs OL}
\end{figure}

\begin{figure}[!tbp]
    \centering
    \includegraphics[width=0.95\linewidth]{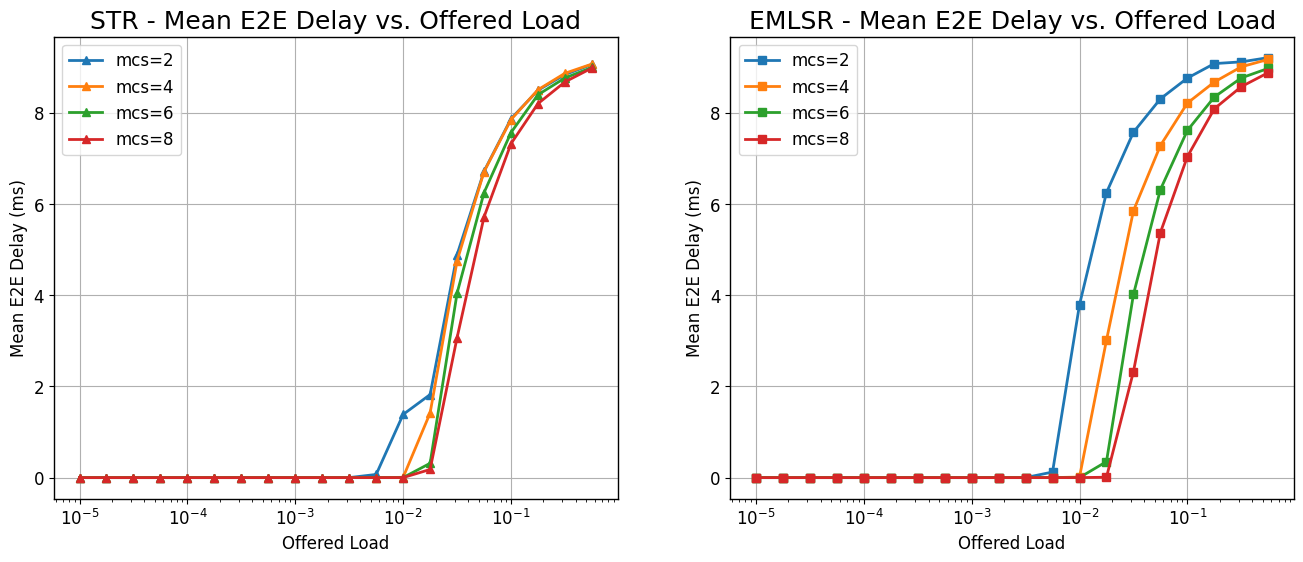}
    \caption{STR vs EMLSR: Mean E2E Delay vs Offered Load with varied MCS}
    \label{fig:Varied MCS - ED vs OL}
\end{figure}

This simulation focuses on comparing STR vs EMLSR performance as the MCS of one of the links is varied. For these simulations, the network consists of one AP with five STAs, operating under either the STR or EMLSR mode, with a payload size of 1500 packets. By default, each link has a 20~MHz channel bandwidth. The offered load was varied from $10^{-5}$ to $10^{-1}$, with a 0.25 step size for the exponent value. Each trace represents a different link 1 MCS value from $\{2, 4, 6, 8\}$. Per \cite{mcs-website}, these MCS values correspond to Modulation types of quadrature phase-shift keying (QPSK), 16-quadrature amplitude modulation (QAM), 64-QAM, and 128-QAM, respectively, each with a Coding type of 3/4. The key aggregate metrics are plotted in Figs. \ref{fig:Varied MCS - T vs OL}-\ref{fig:Varied MCS - ED vs OL}.

Fig.~\ref{fig:Varied MCS - T vs OL} shows the throughput vs offered load as the MCS of each MLD is varied. From this plot, the network whose MLDs employ STR achieve higher saturated throughputs at each MCS value compared to the EMLSR counterpart. For example, when \texttt{mcs=2}, the STR network's throughput saturates around 80~Mbps while the EMLSR network's saturates around 20~Mbps. However, both types of networks observe a similar trend: as the MCS increases by 2, the saturated throughput increases by 20~Mbps. This makes sense as the higher the MCS value, the higher-order the modulation scheme, which allows for increased datarates \cite{mcs-website}.

Meanwhile, Figs. \ref{fig:Varied MCS - QD vs OL}, \ref{fig:Varied MCS - AD vs OL}, and \ref{fig:Varied MCS - ED vs OL} compare the mean queuing, access, and end-to-end delays, respectively, of the STR and EMLSR networks. Since the end-to-end delay is the sum of queuing and access delays, and because queuing delay is orders of magnitude larger than access delay, Figs. \ref{fig:Varied MCS - QD vs OL} and \ref{fig:Varied MCS - ED vs OL} are similar. However, it is still possible to extract the trend that as MCS increases, the saturated network's delay decreases, regardless of operating mode. From Fig. \ref{fig:Varied MCS - AD vs OL}, for applications requiring lower access delay, STR should be favored when \texttt{mcs=\{2,4\}}, while EMLSR should be favored when \texttt{mcs=8}. Both modes appear to have the same access delay when \texttt{mcs=6}.

\subsection{STR vs EMLSR - Varied BW}
\begin{figure}[!tbp]
    \centering
    \includegraphics[width=0.95\linewidth]{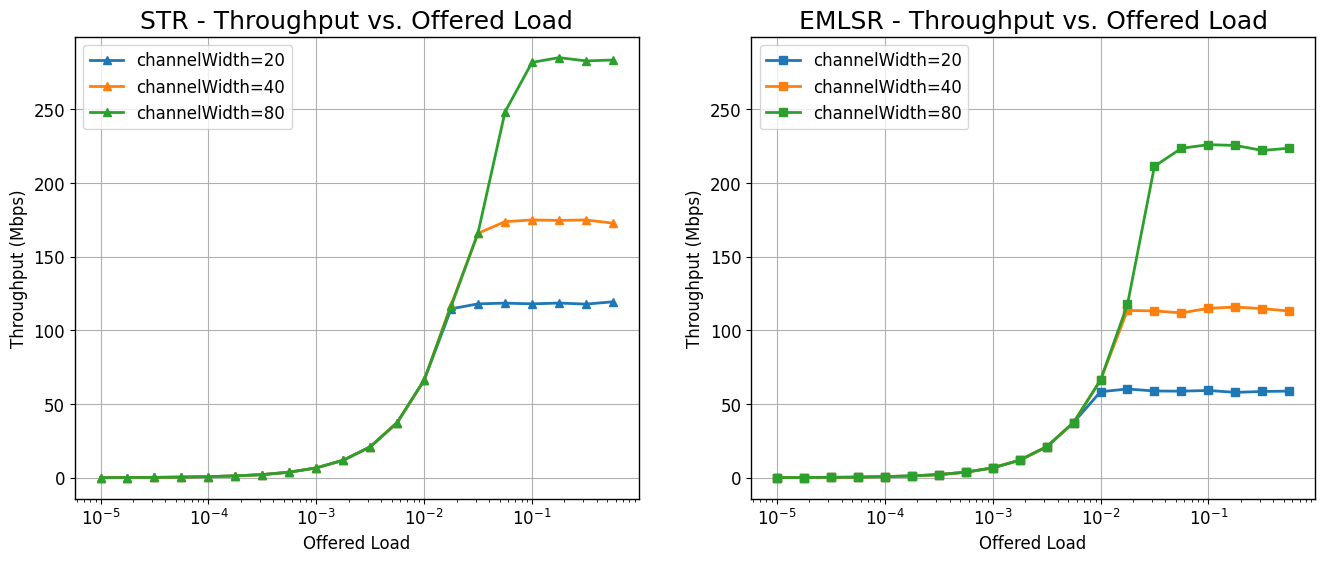}
    \caption{STR vs EMLSR: Throughput vs Offered Load with varied BW}
    \label{fig:Varied BW - T vs OL}
\end{figure}

\begin{figure}[!tbp]
    \centering
    \includegraphics[width=0.95\linewidth]{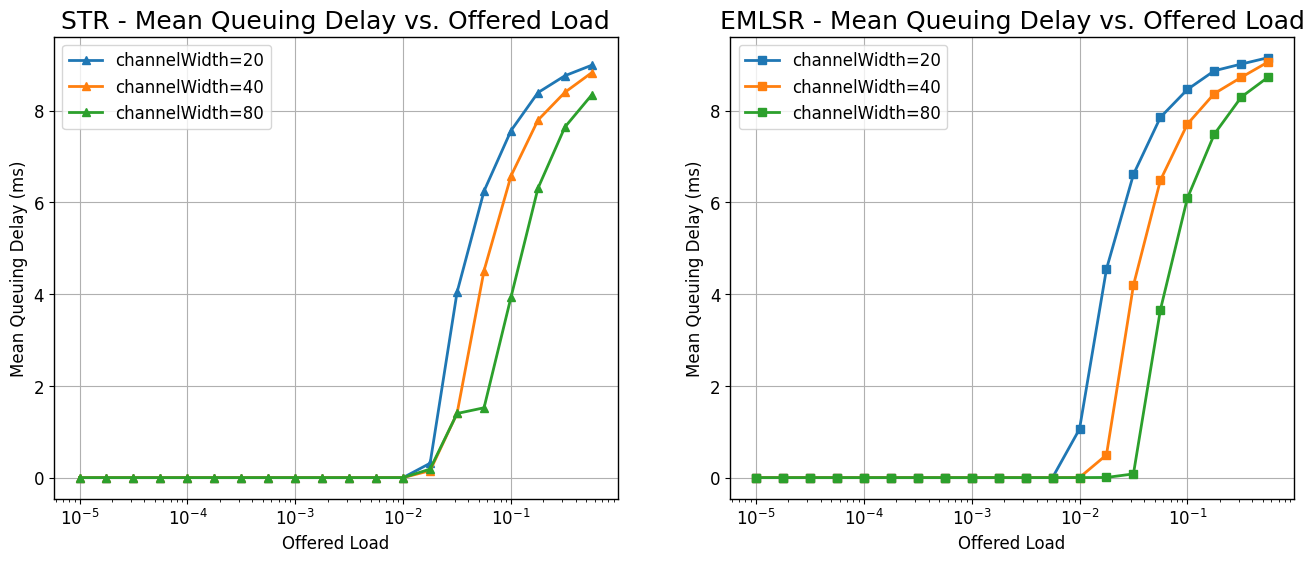}
    \caption{STR vs EMLSR: Mean Queuing Delay vs Offered Load with varied BW}
    \label{fig:Varied BW - QD vs OL}
\end{figure}

\begin{figure}[!tbp]
    \centering
    \includegraphics[width=0.95\linewidth]{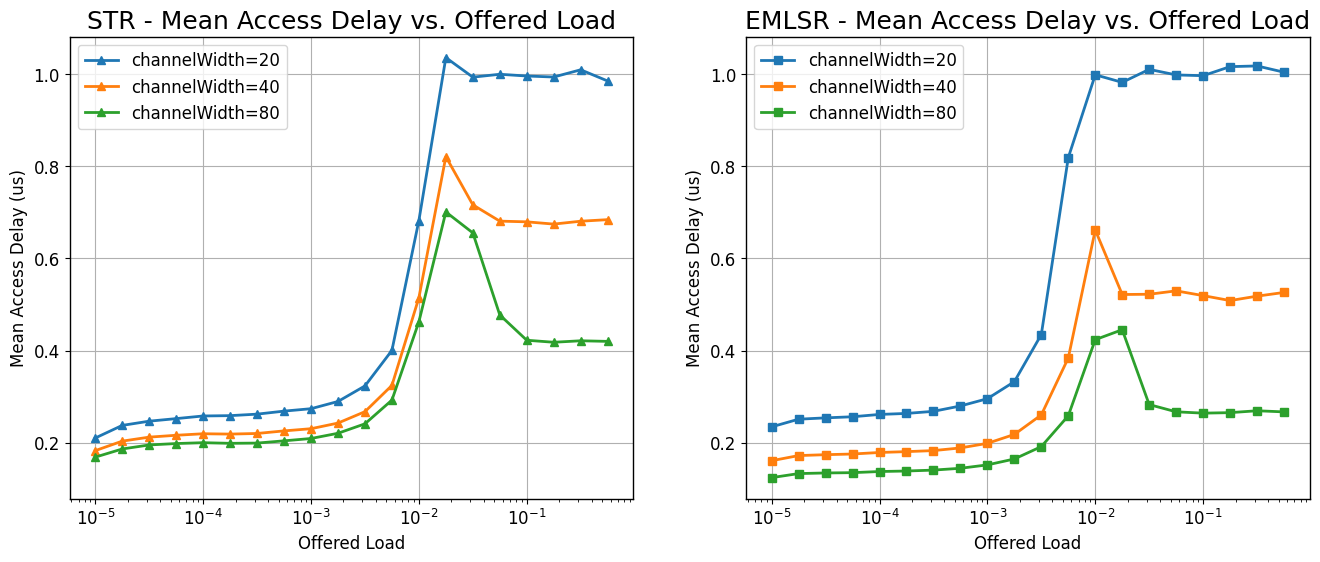}
    \caption{STR vs EMLSR: Mean Access Delay vs Offered Load with varied BW}
    \label{fig:Varied BW - AD vs OL}
\end{figure}

\begin{figure}[!tbp]
    \centering
    \includegraphics[width=0.95\linewidth]{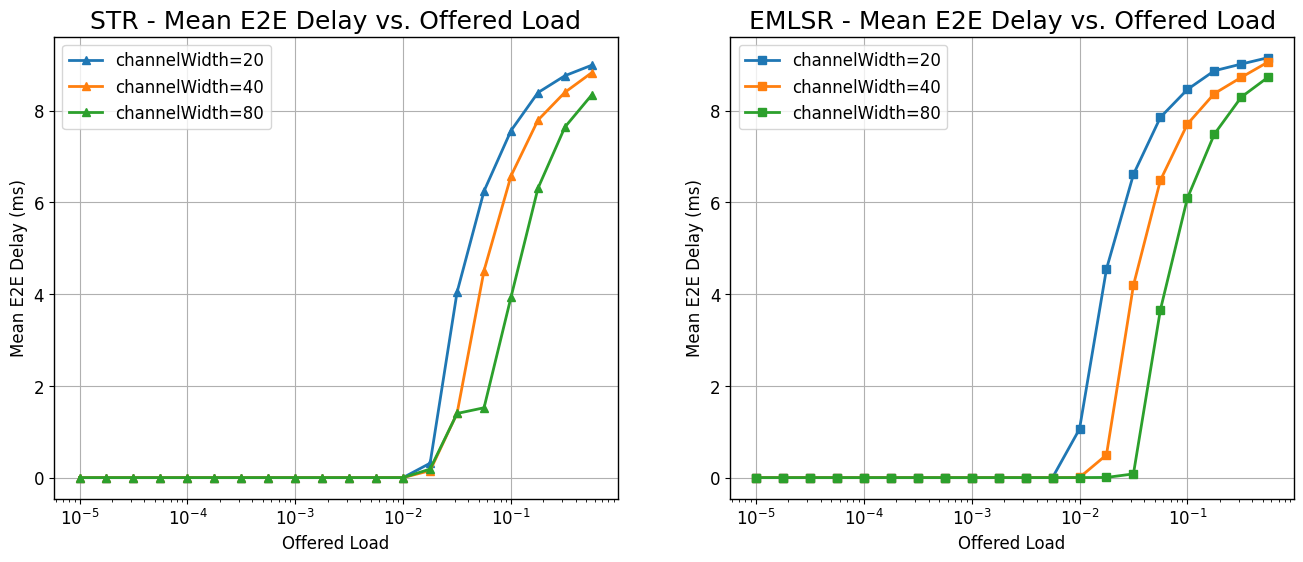}
    \caption{STR vs EMLSR: Mean E2E Delay vs Offered Load with varied BW}
    \label{fig:Varied BW - ED vs OL}
\end{figure}

Similar to the simulation of \ref{subsec:STR vs EMSR - Varied MCS}, this simulation focuses on comparing STR vs EMLSR performance, but this time as the BW of one of the links is varied. As in \ref{subsec:STR vs EMSR - Varied MCS}, the network for this simulation consists of one AP with five STAs, operating under either the STR or EMLSR mode, with a payload size of 1500 packets. By default, each link has an MCS value of 6. The offered load was varied from $10^{-5}$ to $10^{-1}$, with a 0.25 step size for the exponent value. Now, each trace represents a different link 1 BW value from $\{20, 40, 80\}$~MHz. The key aggregate metrics are plotted in Figs. \ref{fig:Varied BW - T vs OL}-\ref{fig:Varied BW - ED vs OL}.

Fig.~\ref{fig:Varied BW - T vs OL} shows the throughput vs offered load as the BW of each MLD is varied. From this plot, at each \texttt{channelWidth} value, the STR network again achieved higher throughput than the EMLSR counterpart. However, for each BW value, the $\lambda$ value which saturates the network is higher for STR compared to EMLSR; this suggests the STR network can handle more traffic than an EMLSR network can. Finally, the trend that increasing BW increases the saturated throughput is observed regardless of operating mode.

Meanwhile, Figs. \ref{fig:Varied BW - QD vs OL}, \ref{fig:Varied BW - AD vs OL}, and \ref{fig:Varied BW - ED vs OL} compare the mean queuing, access, and end-to-end delays, respectively, of the STR and EMLSR networks. As before, since the end-to-end delay is the sum of queuing and access delays, and because queuing delay is orders of magnitude larger than access delay, Figs. \ref{fig:Varied BW - QD vs OL} and \ref{fig:Varied BW - ED vs OL} are similar. However, it is still possible to extract the trend that as BW increases, the saturated network's delay decreases, regardless of operating mode. From Fig. \ref{fig:Varied BW - AD vs OL}, for applications requiring lower access delay, EMLSR should be favored this time when \texttt{channelWidth=\{40,80\}}

\subsection{STR vs EMLSR - Under Interference}
The final series of simulations focuses on comparing STR and EMLSR network performance while SLDs are added to interfere with the network. 

\subsubsection{Scenario 1 - Multi-MLD Network under Asymmetric Interference}
\begin{figure}[!tbp]
    \centering
    \includegraphics[width=0.95\linewidth]{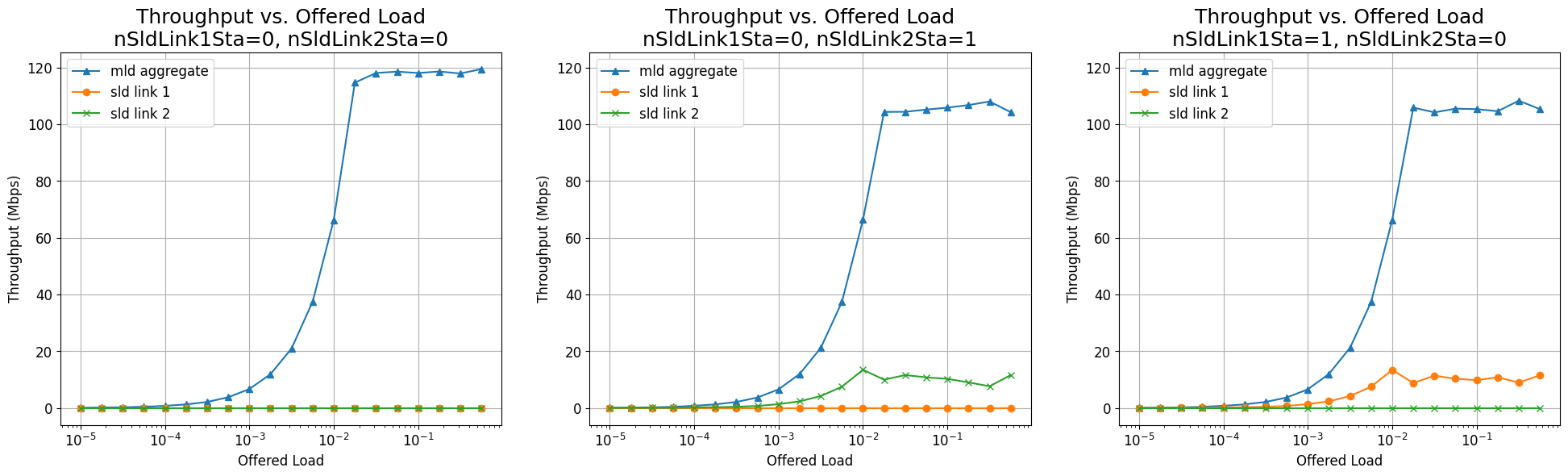}
    \caption{Multi-MLD STR Network: Throughput vs Offered Load under Asymmetric Interference}
    \label{fig:Interference asym STR - T vs OL}
\end{figure}

\begin{figure}[!tbp]
    \centering
    \includegraphics[width=0.95\linewidth]{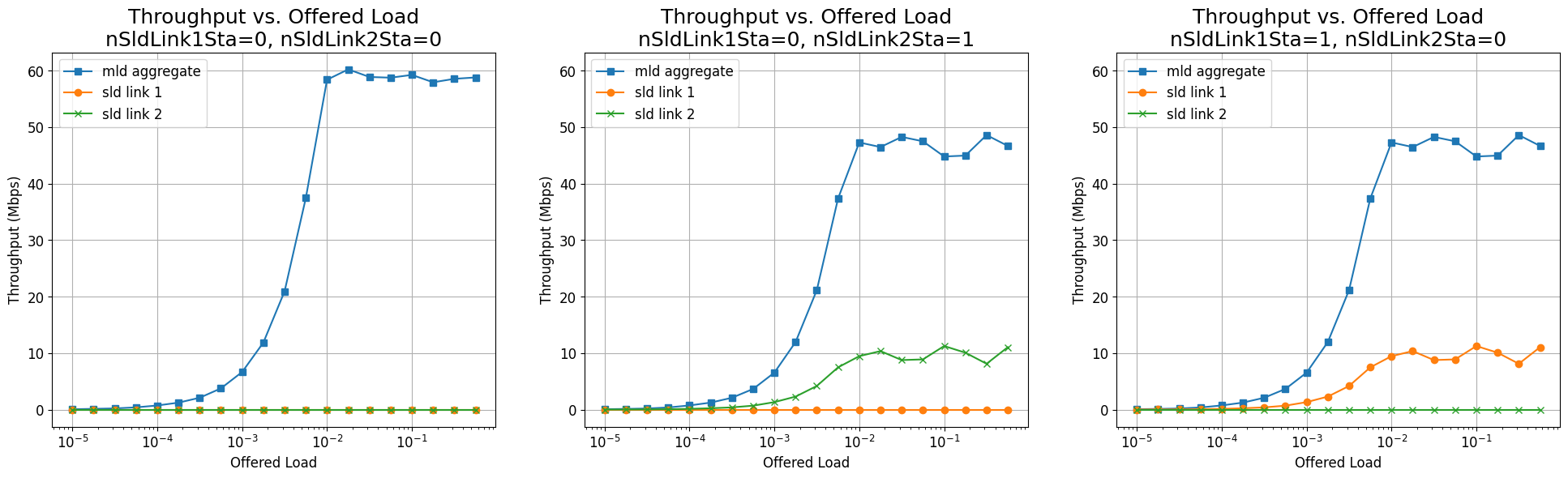}
    \caption{Multi-MLD EMLSR Network: Throughput vs Offered Load under Asymmetric Interference}
    \label{fig:Interference asym EMLSR - T vs OL}
\end{figure}

The first scenario considers a network consisting of one AP with five MLDs, all operating under either STR or EMLSR, with a payload size of 1500 packets. One interfering SLD is then added to one link at a time. The key metrics are plotted in Figs. \ref{fig:Interference asym STR - T vs OL} and \ref{fig:Interference asym EMLSR - T vs OL}.

Regardless of operating mode, the saturated throughput decreases by the same amount of about 10~Mbps when a single interfering SLD is added to either link 1 or link 2. As in previous simulations, the STR network achieves a higher saturated throughput than the EMLSR network does, under this configuration.

\subsubsection{Scenario 2 - Multi-MLD Network under Symmetric Interference}
\begin{figure}[!tbp]
    \centering
    \includegraphics[width=0.95\linewidth]{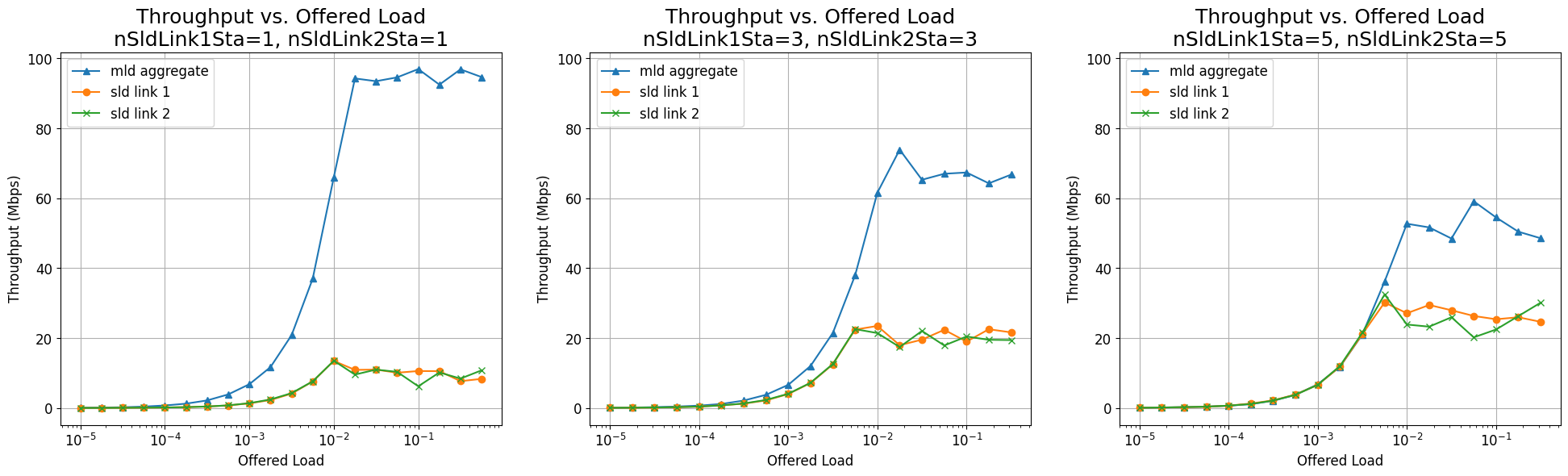}
    \caption{Multi-MLD STR Network: Throughput vs Offered Load under Symmetric Interference}
    \label{fig:Interference sym STR - T vs OL}
\end{figure}

\begin{figure}[!tbp]
    \centering
    \includegraphics[width=0.95\linewidth]{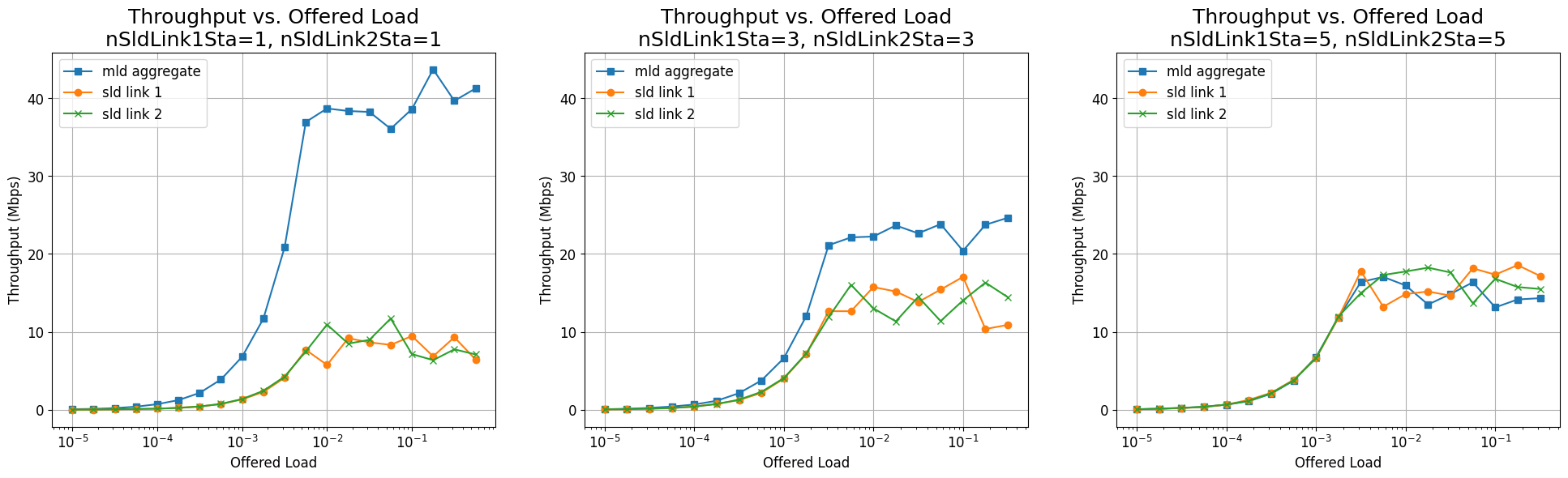}
    \caption{Multi-MLD EMLSR Network: Throughput vs Offered Load under Symmetric Interference}
    \label{fig:Interference sym EMLSR - T vs OL}
\end{figure}

The second scenario again considers a network consisting of one AP with five MLDs, all operating under either STR or EMLSR, with a payload size of 1500 packets. This time, however, the same number of SLDs is added to both links at the same time. The key metrics are plotted in Figs. \ref{fig:Interference sym STR - T vs OL} and \ref{fig:Interference sym EMLSR - T vs OL}.

Regardless of operating mode, the MLD aggregate saturated throughput decreases as more SLDs are added. As more and more SLDs are added, the aggregate MLD throughput and each per-link SLD throughput reach an equilibrium, especially evident in \ref{fig:Interference sym EMLSR - T vs OL}. Once again, the STR network is able to achieve higher saturated throughput at each different number of SLDs compared to EMLSR. 

\subsubsection{Scenario 3 - Single MLD - STR}
The third scenario considers a network consisting of one AP with one MLD, all operating under STR, with a payload size of 1500 packets. The performance is evaluated upon increasing the number of Slds from 5, 10, 15, 20 upon interference due to varying offloads.

\begin{figure}[!tbp]
    \centering
    \includegraphics[width=0.8\linewidth]{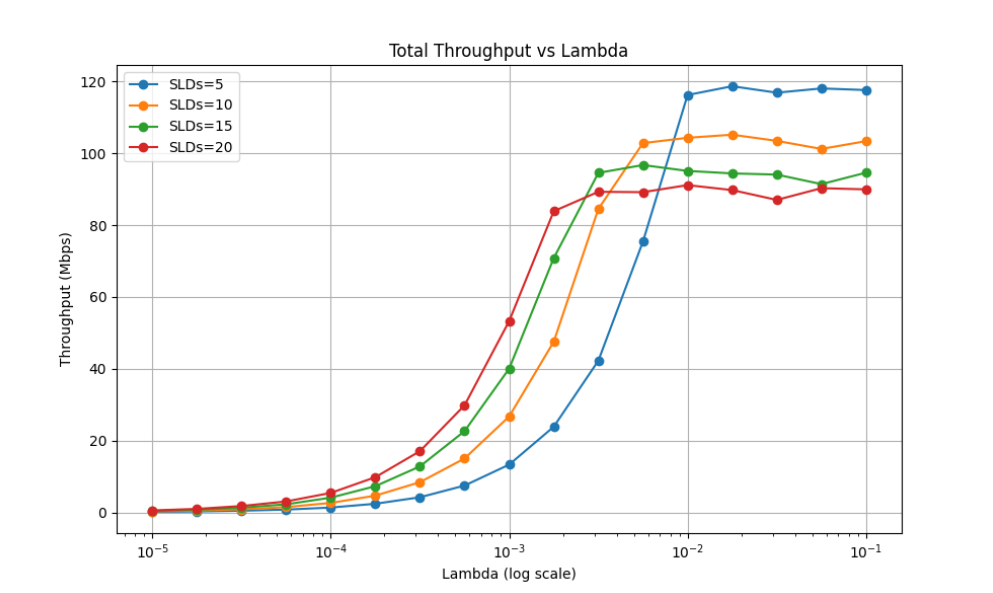}
    \caption{Single-MLD STR Network: Throughput Comparison}
    \label{fig:1MLD_throughput_vs_lambda}
\end{figure}

Fig.~\ref{fig:1MLD_throughput_vs_lambda} shows the throughput variation in STR with one MLD upon increasing number of SLDs. In the previous simulations, it was observed that with five MLDs, the throughput saturates at a lower value of lambda due to contention for resources among multiple MLDs across the links. In comparison, one MLD with varying SLDS performs better due to reduced congestion and overall, it saturates at a higher value of offload. Additionally, it is observed that as the number of SLDs per link increases, a lower throughput is obtained and it saturates as a lower value of lambda. 

\begin{figure}[!tbp]
    \centering
    \includegraphics[width=0.8\linewidth]{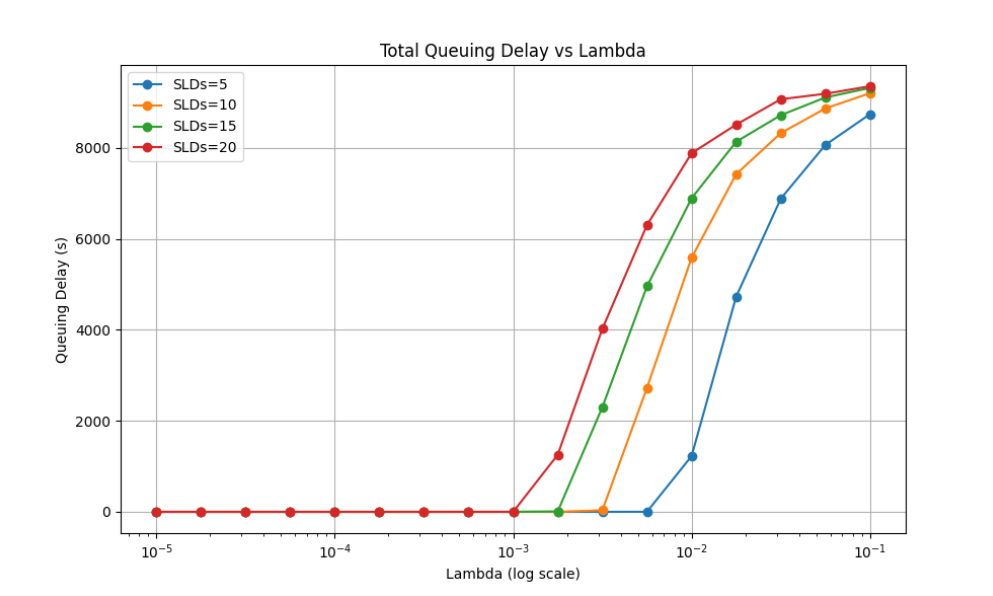}
    \caption{Single-MLD STR Network: Queuing Delay Comparison}
    \label{fig:1MLD_queuingD_vs_lambda}
\end{figure}

\begin{figure}[!tbp]
    \centering
    \includegraphics[width=0.8\linewidth]{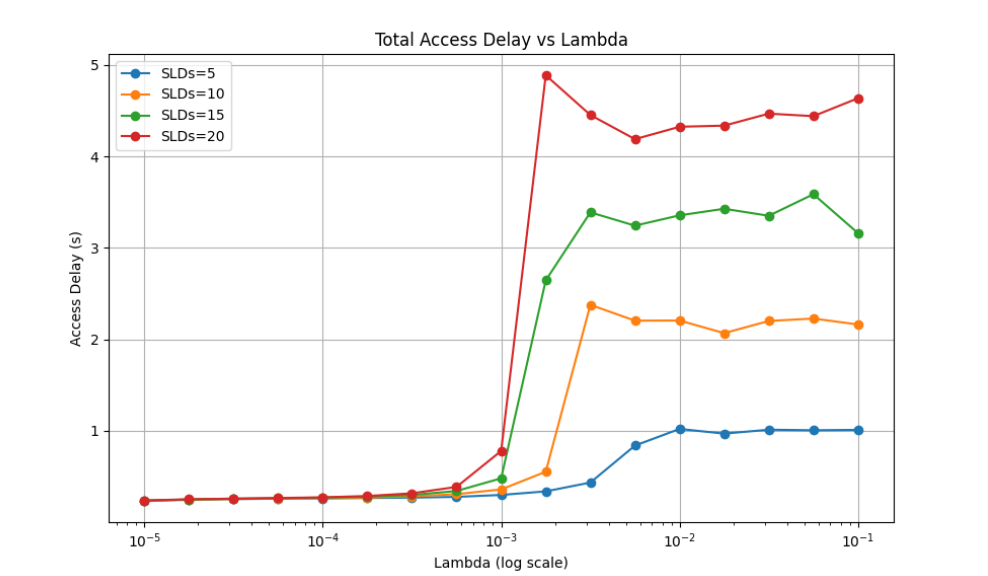}
    \caption{Single-MLD STR Network: Access Delay Comparison}
    \label{fig:1MLD_accessD_vs_lambda}
\end{figure}

\begin{figure}[!tbp]
    \centering
    \includegraphics[width=0.8\linewidth]{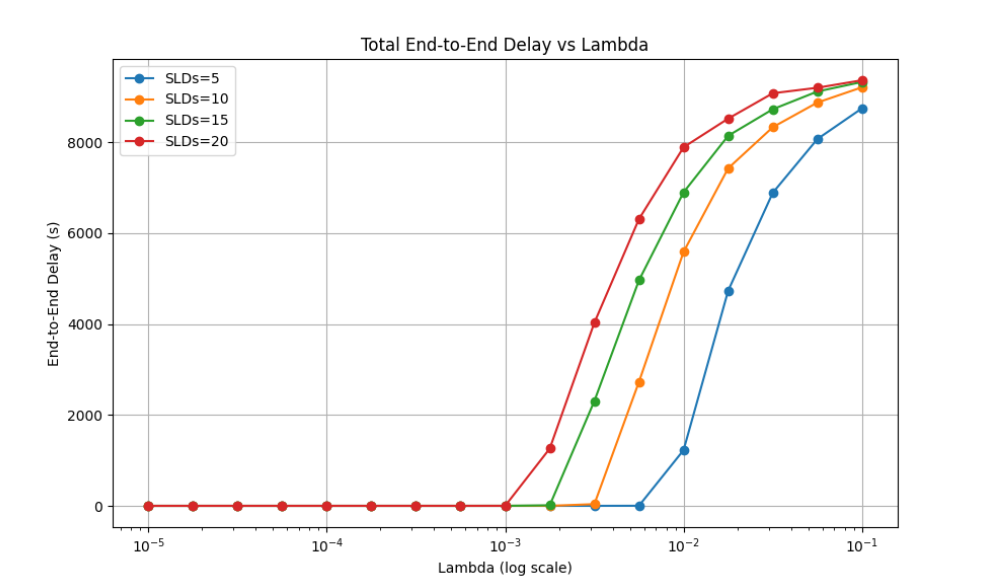}
    \caption{Single-MLD STR Network: End to End Delay Comparison}
    \label{fig:1MLD_e2eD_vs_lambda}
\end{figure}

With five MLDs, latency is higher due to increase in overall network traffic due to multiple MLDs, leading to longer backoff, queuing  and contention for access. Fig.~\ref{fig:1MLD_queuingD_vs_lambda}, Fig.~\ref{fig:1MLD_accessD_vs_lambda}, Fig.~\ref{fig:1MLD_e2eD_vs_lambda} show the latency trends in one MLD with varying SLDs and it is observed that delays experienced are lower since there is to contention for access. However, as the number of SLDs increases per link, the delays sharply increase due to congestion and traffic, yet it happens at a relatively higher offload value.

\subsubsection{Scenario 4 - Single-MLD Network under Symmetric Interference for EMLSR}

The fourth scenario considers utilizing a single MLD with constant lambda, whilst varying the number of SLD's and varying the offload of SLD's. the network consists of 1 AP with a payload size of 1500 packets, and the network is being evaluated by simulated nSldLink1 and nSldLink2 from [5,10,15, 20], but as symmetric points. For this scenario, our lambda was stopped at -2, with a step size of 1.

\begin{figure}[!h]
    \centering
    \includegraphics[width=0.8\linewidth]{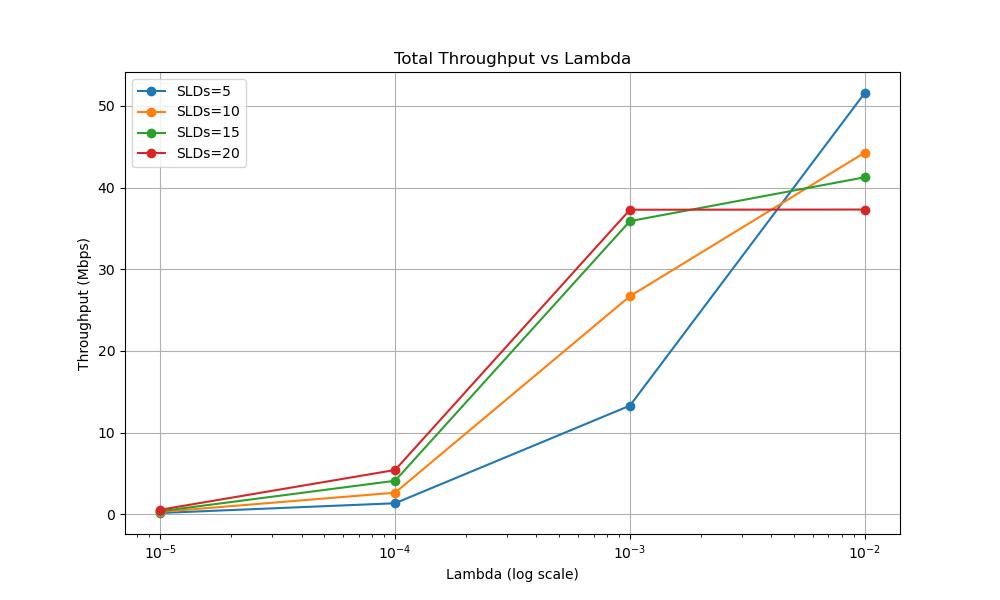}
    \caption{Single-MLD eMLSR Network: Throughput Comparison}
    \label{fig:1MLD_vSLD_throughput}
\end{figure}

\begin{figure}[!h]
    \centering
    \includegraphics[width=0.8\linewidth]{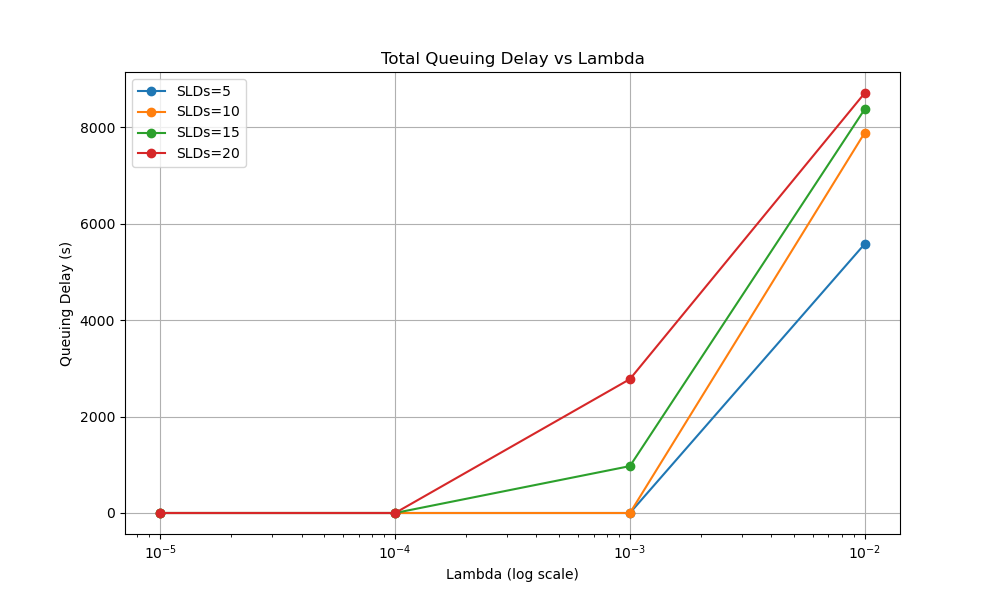}
    \caption{Single-MLD eMLSR Network: Queuing Delay Comparison}
    \label{fig:1MLD_vSLD_que}
\end{figure}

\begin{figure}[!h]
    \centering
    \includegraphics[width=0.8\linewidth]{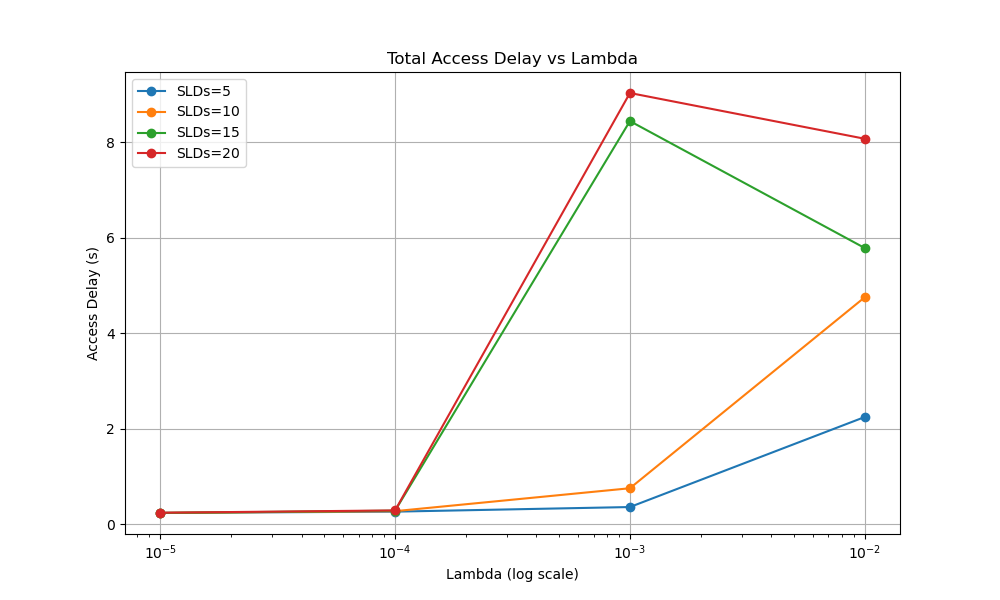}
    \caption{Single-MLD eMLSR Network: Access Comparison}
    \label{fig:1MLD_vSLD_access}
\end{figure}

\begin{figure}[!h]
    \centering
    \includegraphics[width=0.8\linewidth]{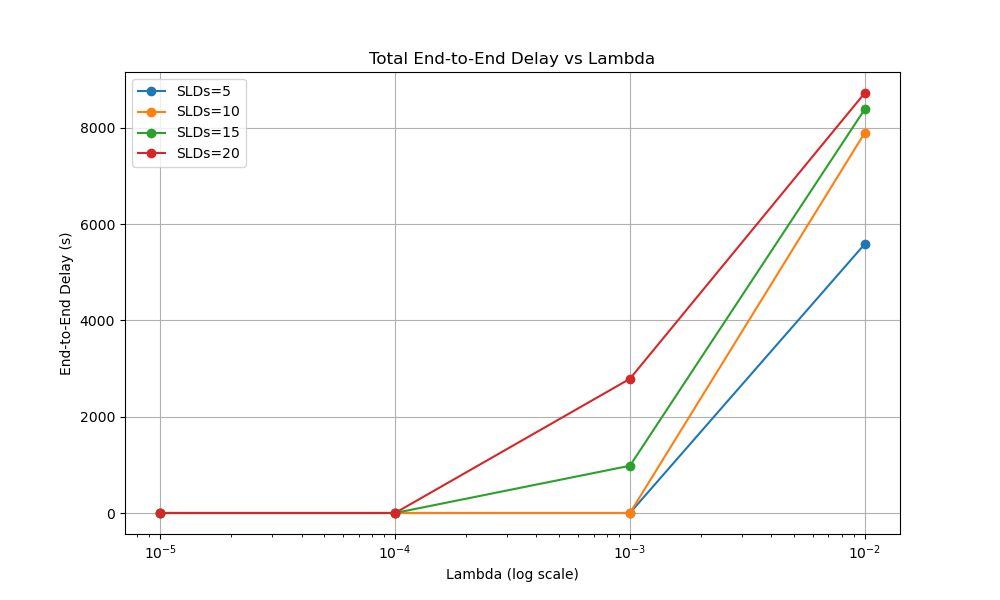}
    \caption{Single-MLD eMLSR Network: End to End Comparison}
    \label{fig:1MLD_vSLD_e2e}
\end{figure}

Fig. ~\ref{fig:1MLD_vSLD_throughput} shows the throughput for all the stations tested, and as our nSTA's increase the saturation of throughput decreases. For example, at nSTa = 20 the throughput saturates at a lower value of lambda. The next plots Fig.~\ref{fig:1MLD_vSLD_que}, Fig.~\ref{fig:1MLD_vSLD_access}, Fig.~\ref{fig:1MLD_vSLD_e2e} show the latency patterns for 1 MLD and varied SLD's. A general pattern for the queueing and end-to-end delay plots is that as the number of SLD increase the rate of delay's sharply increases, this happens at the same offload value for each of the delays. As for the access delay plot, the general trend is that as nSTA's increase the access delay increases too. 

%% file: conclusion.tex
\section{Conclusion and Future Work}
\label{sec:conclusion}

In this study, we investigated the performance of Simultaneous Transmission and Reception (STR) and Enhanced Multi-Link Single Radio (EMLSR) within the context of Wi-Fi 7's Multi-Link Operation (MLO). Through a series of simulations using the ns-3 network simulator, we analyzed how each technique performs under varying conditions, such as changes in modulation and coding schemes (MCS), channel bandwidth, and interference levels. Our results exhibiting the throughput improvements and latency reductions offered by STR and EMLSR compared to SLO verify the benefits of Wi-Fi 7 over older Wi-Fi generations. From our observations and simulation scenarios, STR consistently outperformed EMLSR.

Future work will focus on further refining these simulations to explore additional real-world scenarios and traffic patterns. Additionally, implementing energy efficiency comparisons should identify scenarios in which EMLSR should be favored. Incorporating machine learning (ML) models through the ns3-ai framework~\cite{ns3ai} presents an exciting opportunity to optimize traffic allocation dynamically and improve network performance predictions. By leveraging AI/ML, we can enhance adaptive decision-making for selecting STR or EMLSR modes based on real-time network conditions. Specifically, AI/ML tools could be used to learn the best link allocation strategy to avoid interference \cite{ali2023federatedreinforcementlearningframework}, or to select the least congested channel available \cite{iturriarivera2023channelselectionwifi7}. Additionally, investigating hybrid approaches that combine the strengths of both STR and EMLSR may provide more robust solutions for diverse networking environments.

%% file: main.bbl
\begin{thebibliography}{10}
\providecommand{\url}[1]{#1}
\csname url@samestyle\endcsname
\providecommand{\newblock}{\relax}
\providecommand{\bibinfo}[2]{#2}
\providecommand{\BIBentrySTDinterwordspacing}{\spaceskip=0pt\relax}
\providecommand{\BIBentryALTinterwordstretchfactor}{4}
\providecommand{\BIBentryALTinterwordspacing}{\spaceskip=\fontdimen2\font plus
\BIBentryALTinterwordstretchfactor\fontdimen3\font minus \fontdimen4\font\relax}
\providecommand{\BIBforeignlanguage}[2]{{%
\expandafter\ifx\csname l@#1\endcsname\relax
\typeout{** WARNING: IEEEtran.bst: No hyphenation pattern has been}%
\typeout{** loaded for the language `#1'. Using the pattern for}%
\typeout{** the default language instead.}%
\else
\language=\csname l@#1\endcsname
\fi
#2}}
\providecommand{\BIBdecl}{\relax}
\BIBdecl

\bibitem{10.1145/3659111.3659116}
\BIBentryALTinterwordspacing
M.~Shen, J.~Zhang, H.~Yin, S.~Roy, and Y.~Gao, ``Delay in multi-link operation in ns-3: Validation and impact of traffic splitting,'' in \emph{Proceedings of the 2024 Workshop on Ns-3}, ser. WNS3 '24.\hskip 1em plus 0.5em minus 0.4em\relax New York, NY, USA: Association for Computing Machinery, 2024, p. 19–26. [Online]. Available: \url{https://doi.org/10.1145/3659111.3659116}
\BIBentrySTDinterwordspacing

\bibitem{10624802}
C.-L. Tai, M.~Eisen, D.~Akhmetov, D.~Das, D.~Cavalcanti, and R.~Sivakumar, ``Model-free dynamic traffic steering for multi-link operation in ieee 802.11be,'' in \emph{2024 IEEE International Conference on Machine Learning for Communication and Networking (ICMLCN)}, 2024, pp. 44--49.

\bibitem{10495351}
L.~Zhang, H.~Yin, S.~Roy, L.~Cao, X.~Gao, and V.~Sathya, ``Ieee 802.11be network throughput optimization with multilink operation and ap controller,'' \emph{IEEE Internet of Things Journal}, vol.~11, no.~13, pp. 23\,850--23\,861, 2024.

\bibitem{https://doi.org/10.1155/2022/7018360}
\BIBentryALTinterwordspacing
X.~Lan, X.~Zu, and J.~Yang, ``Enhanced multilink single-radio operation for the next-generation ieee 802.11 be wi-fi systems,'' \emph{Security and Communication Networks}, vol. 2022, no.~1, p. 7018360, 2022. [Online]. Available: \url{https://onlinelibrary.wiley.com/doi/abs/10.1155/2022/7018360}
\BIBentrySTDinterwordspacing

\bibitem{v3.41-release}
\BIBentryALTinterwordspacing
nsnam. ns-3.41. [Online]. Available: \url{https://www.nsnam.org/releases/ns-3-41/}
\BIBentrySTDinterwordspacing

\bibitem{group5-github}
\BIBentryALTinterwordspacing
G.~5. ee595-f24-group5. [Online]. Available: \url{https://github.com/kevinho-uw/ee595-f24-group5}
\BIBentrySTDinterwordspacing

\bibitem{alsakati2023performance80211bewifi7}
\BIBentryALTinterwordspacing
M.~Alsakati, C.~Pettersson, S.~Max, V.~N. Moothedath, and J.~Gross, ``Performance of 802.11be wi-fi 7 with multi-link operation on ar applications,'' 2023. [Online]. Available: \url{https://arxiv.org/abs/2304.01693}
\BIBentrySTDinterwordspacing

\bibitem{10149044}
M.~Carrascosa-Zamacois, G.~Geraci, E.~Knightly, and B.~Bellalta, ``Wi-fi multi-link operation: An experimental study of latency and throughput,'' \emph{IEEE/ACM Transactions on Networking}, vol.~32, no.~1, pp. 308--322, 2024.

\bibitem{mcs-website}
\BIBentryALTinterwordspacing
S.~Networks. Mcs index, modulation and coding scheme. [Online]. Available: \url{https://mcsindex.net/}
\BIBentrySTDinterwordspacing

\bibitem{ns3ai}
\BIBentryALTinterwordspacing
H.~Yin, P.~Liu, K.~Liu, L.~Cao, L.~Zhang, Y.~Gao, and X.~Hei, ``Ns3-ai: Fostering artificial intelligence algorithms for networking research,'' in \emph{Proceedings of the 2020 Workshop on Ns-3}, ser. WNS3 2020.\hskip 1em plus 0.5em minus 0.4em\relax New York, NY, USA: Association for Computing Machinery, 2020, p. 57–64. [Online]. Available: \url{https://doi.org/10.1145/3389400.3389404}
\BIBentrySTDinterwordspacing

\bibitem{ali2023federatedreinforcementlearningframework}
\BIBentryALTinterwordspacing
R.~Ali and B.~Bellalta, ``A federated reinforcement learning framework for link activation in multi-link wi-fi networks,'' 2023. [Online]. Available: \url{https://arxiv.org/abs/2304.14720}
\BIBentrySTDinterwordspacing

\bibitem{iturriarivera2023channelselectionwifi7}
\BIBentryALTinterwordspacing
P.~E. Iturria-Rivera, M.~Chenier, B.~Herscovici, B.~Kantarci, and M.~Erol-Kantarci, ``Channel selection for wi-fi 7 multi-link operation via optimistic-weighted vdn and parallel transfer reinforcement learning,'' 2023. [Online]. Available: \url{https://arxiv.org/abs/2307.05419}
\BIBentrySTDinterwordspacing

\end{thebibliography}
